# The Positive Experience Principle: Forecasting Conscious Choices with AI Embeddings

Zheng Su[1]*, Mingyan Fang[2]*

* Correspondence to suzheng@whu.edu.cn, fangmingyan@aliyun.com

## Abstract

A fundamental challenge in the science of consciousness is the lack of a universal, predictive framework for motivated behavior. While existing theories excel at describing specific mechanisms, from neural pathways to computational models, they do not provide a foundational principle that explains the consistent direction of conscious systems toward certain states and away from others. To address this gap, we propose the Positive Experience Principle (PEP), a unifying principle stating that conscious systems have an inherent tendency to move toward states of higher positive subjective experience. This tendency is quantified by a Positive Experience Value (PEV), a scalar metric derived from the physical configurations defined by our earlier Universal Consciousness Code (UCC) theory. The PEP bridges physics, neuroscience, and psychology by positing that diverse behaviors are manifestations of a single, fundamental drive to optimize PEV. The PEP generates testable predictions for the dynamics of conscious systems, offering a path toward a unified science of behavior.

## 1. Introduction

The pursuit of understanding what drives conscious behavior has deep roots in both philosophy and science. Classical hedonistic theories, from Epicurus to Jeremy Bentham's utilitarian calculus, proposed that human action is fundamentally oriented toward maximizing pleasure and minimizing pain [1]. This intuition has found expression across diverse fields: psychological behaviorism demonstrated how organisms learn through reward and punishment [2, 3], while modern neuroscience has mapped the neural substrates of reward processing through dopamine pathways [4, 5].

Contemporary research has significantly advanced our understanding of the mechanisms underlying conscious experience and motivated behavior. Integrated Information Theory (IIT), initially proposed by Giulio Tononi in 2004, attempts to quantify consciousness through mathematical frameworks [6, 7]. Global Workspace Theory (GWT), first introduced in 1988 by cognitive scientist Bernard Baars, explains how information becomes globally accessible in conscious experience [8, 9]. Predictive processing frameworks suggest that the brain constantly generates predictions about sensory input to minimize prediction error [10, 11].

Reinforcement learning theory has provided mathematical models for how organisms learn to maximize rewards [12], while neuroscience has mapped specific reward pathways involving dopamine, serotonin, and other neurotransmitter systems [13]. Approach-avoidance theories have formalized how organisms navigate toward positive and away from negative stimuli [14]. Decision-making research has identified how humans evaluate options through processes resembling expected utility calculations [15, 16].

1 Computational Biology Group, FAB Tech, Yantian, Shenzhen, Guangdong, 518083, China
2 AnyHelix Ltd., The Cloud, Tai Kok Tsui, Hong Kong, China

Despite these advances, a fundamental gap remains: current theories are largely descriptive rather than predictive, focusing on specific mechanisms without offering a universal principle that governs the direction of conscious behavior. For instance, reinforcement learning describes how associations are formed but does not specify what makes an outcome rewarding in the first place. Approach-avoidance theories describe behavioral tendencies without explaining their ultimate origin. Even comprehensive frameworks like IIT focus on the generation of conscious experience rather than its intrinsic motivational dynamics. Consequently, the field lacks a unifying law that can predict behavior, akin to how fundamental principles in physics (e.g., energy minimization) predict the behavior of physical systems.

Our previous work aimed to provide a foundational framework through the Universal Consciousness Code (UCC) theory [17]. This theory posits that consciousness is an intrinsic property of specific configurations of particles ("consciousons"), with each configuration corresponding to a unique subjective experience via a universal code. The UCC provides a static map between physical states and subjective experience. However, it leaves a critical question unanswered: what determines the direction of change within a conscious system?

To address this limitation and introduce a predictive law, we now build upon the UCC framework to propose a fundamental directional principle for conscious systems: the Positive Experience Principle (PEP). We propose that the dynamics of any conscious system are directed toward states of higher positive subjective experience. This experience is quantified by a Positive Experience Value (PEV), a scalar value inherent to each UCC configuration. Just as physical systems follow fundamental gradients (e.g., entropy increase, energy minimization), we propose that conscious systems are governed by the PEP—a principle as fundamental as these physical laws.

The PEP is not only a psychological observation but a proposed law governing the behavior of consciousons. It provides a unified, physical basis for understanding all motivated behavior, from simple reflexes to complex planning.

## 2. The PEP and The PEV Metric

### 2.1. The core postulate

The core postulate of the PEP is that a conscious system will tend to transition to a state that offers more positive experience, i.e. higher Positive Experience Value (PEV).

### 2.2. Simplified model: binary choice situations

To establish the foundational rule, let's consider a highly simplified, isolated conscious system that can only transit to two possible states, $S_1$ and $S_2$, each with an associated PEV, such as a simple binary decision like "pull hand away from hot surface" or "don't pull away." In this idealized scenario, with no constraints or competing drives, the system's choice is governed by a simple rule: it will tend to transition to the state with the higher PEV. The fundamental rule is straightforward:

The probability of transitioning to state $S_2$ from state $S_1$ increases monotonically with the difference in their Positive Experience Value:

$$P(S_1 \rightarrow S_2) \propto f\big(PEV(S_2) - PEV(S_1)\big)$$

where $f$ is a strictly increasing function. The core principle is that larger positive experience differences drive proportionally higher transition probabilities.

For example, if touching a hot stove generates PEV = -20 (intense negative experience) while pulling away generates PEV = +5 (relief and safety), the system will overwhelmingly choose to pull away. This simple rule explains reflexive and instinctive behaviors where only two clear options exist.

In simpler terms, a conscious system will act to maintain or increase its PEV value. It is driven from less preferable states (low PEV) to more preferable states (high PEV).

### 2.3. The Positive Experience Value

The Positive Experience Value (PEV) is the core metric of our principle. It's a single scalar value that quantifies the subjective valence of a conscious state, derived directly from the UCC of the consciousons within that system. A specific UCC configuration corresponds not only to a particular perception (e.g., seeing a bird, feeling warmth) but also to a specific value of positive or negative experience. This value is an inherent property of the UCC itself. For example, the UCC for "feeling love" would have a high PEV, while the UCC for "feeling pain" would have a very low, or negative, PEV.

The PEV is a theoretical metric, a single scalar value P that is a function of a system's UCC:

PEV = F(UCC)

Where F is a mathematical function that maps the vast state space of possible UCCs to a spectrum of subjective valence. A high PEV value corresponds to UCCs for experiences of pleasure, comfort, safety, happiness, and fulfillment. A low PEV value corresponds to UCCs for pain, distress, fear, and sadness. This implies that the UCCs for all conscious experiences can be mapped onto a spectrum from highly negative to highly positive, with neutral experiences residing in the middle.

### 2.4. Complex model: multiple states and future planning

Real conscious systems rarely face simple binary states. Instead, they choose from a set of potential states, where each action can lead to a distribution of possible future states with associated uncertainties.

To model this, consciousness optimizes the Expected Positive Experience Value (expected-PEV). The expected-PEV for a given action is calculated as the sum of the PEV values of each potential outcome state, weighted by the system's perceived intensity of that outcome:

$$\text{expected-PEV(action)} = \sum_{\mathrm{i}} [\mathrm{I_i} \times \mathrm{PEV}(\mathrm{S_i})]$$

Where:

- $S_i$ represents a possible future state resulting from the action.
- $\mathrm{PEV}(S_i)$ is the Positive Experience Value of state $S_i$.

- $I_i$ is the perceived intensity (0 ≤ I ≤ 1) of outcome $S_i$, representing how strongly the system anticipates or weights that particular outcome in its decision-making process.

The perceived intensity may be influenced by different factors such as:

- Perceived probability: How likely the system believes the outcome is to occur
- Temporal proximity: How soon the outcome is expected (near outcomes often feel more intense)
- Vividness: How easily the system can imagine or simulate the outcome

The probability of the system choosing a particular action is then proportional to a function of its expected-PEV:

P(action) ∝ g( expected-PEV(action) )

where g is a strictly increasing function. This framework formalizes the process of planning and deliberation, where the system forecasts the positive experience potential of its available actions before committing to one.

### 2.5. A PEV calculation example: the Diet Dilemma

To illustrate how PEV can be applied in a practical scenario, we first define a relatable everyday decision-making context. Consider the Diet Dilemma: Your brain is running a silent evaluation: "What will lead to maximize my Positive Experience Value (PEV)?" The decision to eat a piece of chocolate cake while on a diet is a complex battle of competing future states.

Your brain doesn't just see "cake." It rapidly evaluates potential associated future states, perceiving each a PEV Value (how good it feels) and a Perceived Intensity (how strongly that outcome weighs on the decision). The action with the highest total Expected PEV wins.

Table 1. Expected Positive Experience Value (PEV) calculation for the diet dilemma

| Potential future state | PEV value | Perceived intensity | Weighted PEV |
|---|---|---|---|
| 1. Taste pleasure | 90 | 1.0 | 90 |
| 2. Guilt & regret | -70 | 0.8 | -56 |
| 3. Weight gain fear | -50 | 0.3 | -15 |
| 4. Social joy | 40 | 0.2 | 8 |
| 5. Philosophical justification | 60 | 0.9 | 54 |
| 6. Social media validation | 30 | 0.5 | 15 |
| Total expected PEV (eating) | | | 96 |
| Total expected PEV (not eating) | | | 0 |

The decision: The sum of the Weighted PEV for eating the cake is +96. The sum for not eating is +0. The action "eat the cake" has a much higher Expected PEV. The immediate pleasure (+90), combined with the powerful philosophical justification (+54) and the potential for social validation (+15), overwhelmingly outweigh the future guilt and fear. The PEP strongly favors this action, and you eat the cake with a sense of justified joy.

### 2.6. Derivation and prediction

The UCC theory proposes that AI embeddings are proxies for UCCs. Therefore, the function F mapping UCC to PEV can potentially be learned by AI systems. By training models on human behavioral data (e.g., choices, preferences, self-reports of well-being) correlated with AI embeddings (proxies for UCC), we can infer the PEV function. This would allow us to predict the behavior of a conscious system by calculating which future action or state has the highest expected PEV.

## 3. Proof-of-Concept Study

### 3.1. Study design

To demonstrate the feasibility of the PEP framework, we conducted a proof-of-concept study testing whether the mapping from experience representations to PEV could be learned from human choice behavior. According to PEP theory, when people choose between experiential alternatives, they preferentially select the option with higher PEV. If this principle holds, and if AI embeddings serve as reasonable proxies for UCC configurations, then we should be able to train a model to predict human choices by learning which experience representations correspond to higher PEV.

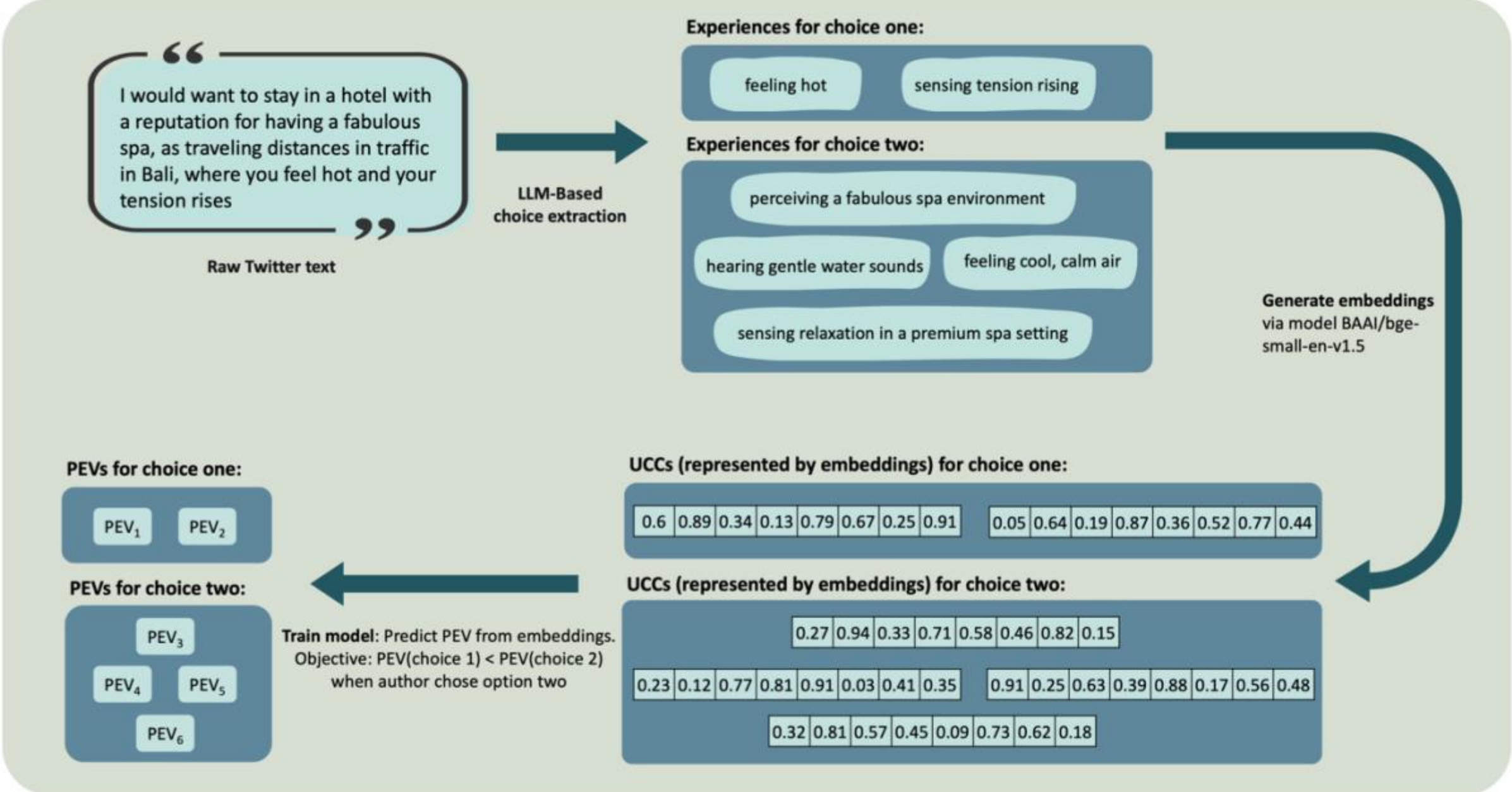

**Figure 1.** Computational framework for learning the UCC-to-PEV mapping. Twitter text describing experiential choices is processed to extract specific experiences via large language model, converted to embeddings (UCC proxies), and used to train a linear model predicting Positive Experience Values (PEV). The model learns that chosen experiences have higher PEV than unchosen alternatives, supporting the Positive Experience Principle.

We collected naturally occurring choice data from Twitter, where users describe decisions between experiential alternatives. From a dataset of over 416,809 tweets, we identified 1,187 cases where someone clearly chose between two distinct action options. For example, a person might describe choosing to "stay in a hotel with a reputation for having a fabulous spa" over "traveling distances in traffic in Bali where you feel hot and your tension rises." Using large language models, we extracted the specific sensory and perceptual experiences associated with each option. For the chosen option (hotel with spa), we identified experiences like "perceiving a fabulous spa environment," "sensing relaxation in a premium spa setting," "feeling cool, calm air," and "hearing gentle water sounds." For the unchosen option (traveling in traffic), we extracted "feeling hot" and "sensing tension rising." We then represented these experiences using AI embeddings, which served as proxies for their UCC configurations.

The core idea was simple: train a model that takes an experience embedding as input and outputs a scalar PEV value. If PEP is correct, the model should learn to assign higher PEV to chosen experiences than to unchosen alternatives. We used a straightforward linear model with a pairwise ranking objective that directly optimizes the constraint PEV(chosen) > PEV(unchosen).

### 3.2. Study results

The model achieved consistent above-chance performance across all validation tests. When presented with a pair of experiences where we knew which one a person actually chose, the model correctly assigned higher PEV to the chosen experience 66.1% of the time, compared to the 50% expected by chance alone (Figure 2). This difference was highly statistically significant ($p \leq 0.0023$, t-test).

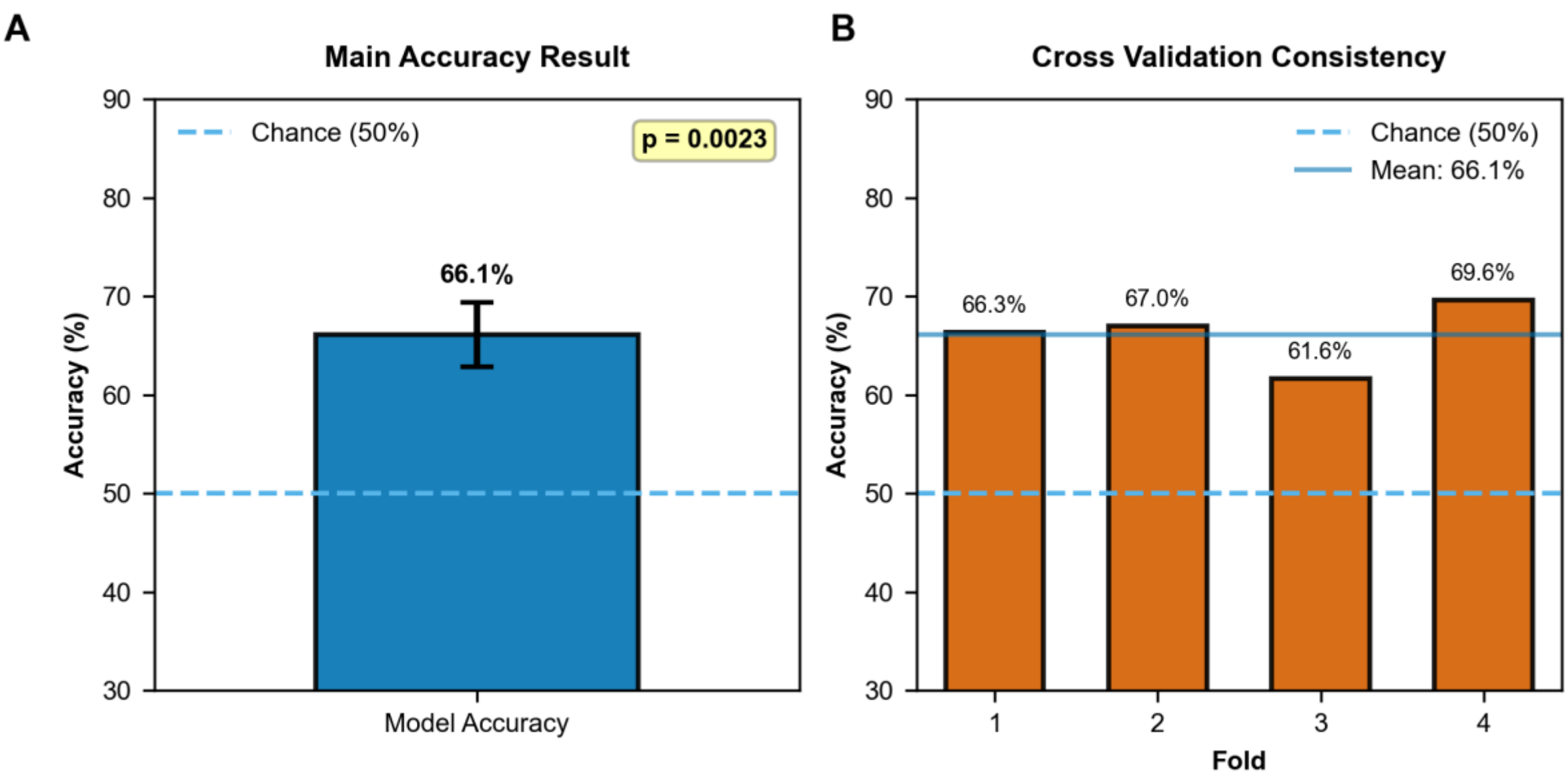

**Figure 2.** Core performance metrics. Panel A shows overall model accuracy summarized across cross-validation folds with 95% confidence intervals; a dashed horizontal line denotes the 50% chance level, and the p-value is reported from a one-sample t-test against chance. Panel B displays accuracy for each validation fold; the dashed line indicates 50% chance and the solid line marks the mean across folds.

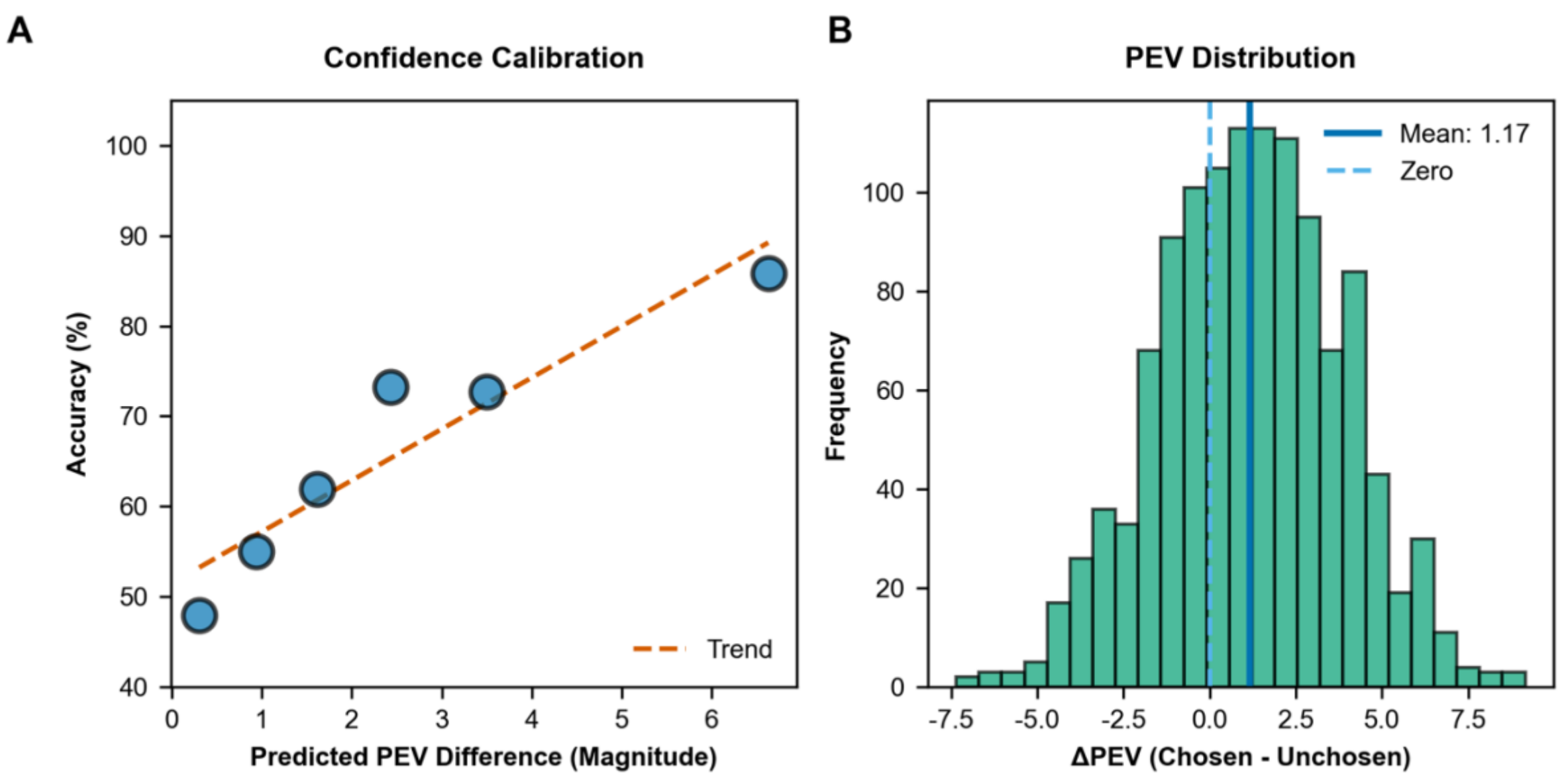

**Figure 3.** Model calibration and PEV differences. Panel A depicts prediction accuracy as a function of the magnitude of the predicted ΔPEV (|PEV_chosen − PEV_unchosen|), using binned estimates and a least-squares trend line to visualize calibration; the x-axis is predicted ΔPEV magnitude and the y-axis is accuracy (%). Panel B presents the distribution of ΔPEV (PEV_chosen − PEV_unchosen) across validation samples; a solid vertical line marks the mean ΔPEV and a dashed vertical line marks zero, with the x-axis as ΔPEV and the y-axis as frequency.

Figure 2B shows the model's performance across four independent validation tests. The consistency of results across different subsets of data suggests that the relationship between experience representations and choice behavior is systematic rather than spurious. The model also demonstrated well-calibrated confidence: when it predicted a large PEV difference between two options, people were more likely to have chosen the higher-PEV option (Figure 3A).

Several additional findings support the core predictions of PEP. The average difference in PEV between chosen and unchosen experiences was substantial (approximately 1.17 on our scale, Figure 3B), indicating that people do not make arbitrary choices but systematically select experiences with meaningfully higher predicted positive value. The model learned this mapping using only a simple linear transformation, suggesting that PEV may be a relatively straightforward projection from high-dimensional experience representations, analogous to how physical quantities like energy summarize complex system states.

## 4. The PEP as a Universal Driver of Behavior

The power of the PEP lies in its universality. It is not a human-specific psychological concept but a principle applicable to any system with UCC-defined states.

Biological drives (animals & humans): Eating, drinking, and seeking warmth are not primary goals. The primary goal is PEV maximization. These activities are highly effective strategies to rapidly increase PEV by resolving homeostatic deficits (hunger, thirst, cold) which create low-PEV UCC states.

Movement and exploration: Locomotion often favors environments yielding comfort or novelty (e.g., migrating to warmer climates). PEP frames this as ascending PEV gradients, where dopamine's role in curiosity and reward anticipation propels exploratory movements to test UCC configurations for higher positive experience.

Social and creative behaviors: Conversation, dancing, creating art, playing music, and social bonding are all pursuits that generate high-PEV UCC configurations. We seek them out because they are powerful drivers of PEV.

Learning and adaptation: Learning is the process of building a model of the world that identifies which actions are most likely to lead to high-PEV states and which are likely to lead to low-PEV states.

Entertainment, singing, dancing: These activities directly boost PEV via rhythmic, aesthetic, or narrative rewards. PEP posits that cultural evolution favors UCC patterns amplifying such positive experiences, explaining the universality of art and play as PEV maximizers.

Evolutionary fitness: From a PEP perspective, behaviors that enhance survival and reproduction are evolutionarily selected for precisely because they are strongly correlated with high PEV (e.g., the pleasure of eating, bonding, mating). PEV can be seen as the subjective, experiential correlate of evolutionary fitness.

Maladaptive behaviors: The PEP also explains maladaptive patterns. Addiction occurs because a substance or activity directly and powerfully hijacks the brain's UCC, generating an artificially high PEV value, compelling the system to pursue it above all other, more sustainable strategies.

## 5. Discussion

We have introduced the Positive Experience Principle (PEP), a proposed universal principle stating that conscious systems evolve toward states with higher Positive Experience Value (PEV). By extending the Universal Consciousness Code (UCC) theory from a static mapping of physical states to subjective experiences into a dynamical framework, the PEP provides a missing element: a predictive principle governing the direction of conscious change. This formulation advances the study of consciousness beyond descriptive accounts, offering a candidate for a mathematically grounded and physically interpretable law of conscious dynamics.

The central contribution of this work is to formalize the link between subjective experience and behavioral direction through three key advances: First, the PEP reframes the long-standing pleasure-seeking principle as a law rooted in consciousness itself, rather than treating it as an empirical observation about behavior. Second, it introduces the PEV as a scalar metric that can, in principle, be derived from the UCC state of a system, transforming qualitative concepts of "value" or "pleasure" into quantifiable properties of physical configurations. Third, it offers a unified framework that posits diverse drives, whether for food, safety, social connection, or achievement, as emergent manifestations of a single underlying imperative to optimize positive subjective experience.

The proof-of-concept study results provide preliminary empirical support for the Positive Experience Principle. The fact that a computational model can learn to predict human choices from experience representations demonstrates three key aspects of the PEP framework:

First, experience representations contain extractable information about subjective value. The AI embeddings we used captured semantic content relevant to positive experience, allowing the model to distinguish between more and less preferable states.

Second, the learned PEV values show the predicted directionality: chosen experiences generally received higher PEV assignments than unchosen alternatives. This aligns with PEP's central claim that conscious systems transition toward states of higher positive subjective experience.

Third, the simplicity of the successful model supports PEP's formulation of behavior as fundamentally driven by a scalar metric. Just as physical systems follow energy gradients, our results suggest that experiential choices may follow PEV gradients in the space of possible conscious states.

The proof-of-concept study serves as a demonstration rather than a comprehensive validation of PEP. Several important limitations should be noted. Our data came from written descriptions of choices on social media, which may not fully capture the richness of in-the-moment experiential states. The AI embeddings we used are derived from linguistic descriptions rather than direct measurements of neural activity, making them imperfect proxies for UCC configurations. Additionally, our model learned relative PEV differences rather than absolute values, which is sufficient for predicting binary choices but limits broader interpretation.

Future work should address these limitations through multiple approaches. Larger and more diverse datasets spanning different choice contexts and modalities would strengthen the evidence. Experimental manipulations that systematically vary experience descriptions in controlled settings could establish causal relationships rather than correlations. Multi-modal embeddings combining vision, audio, and physiological signals could capture experiential states more completely. Finally, longitudinal studies could track how the UCC-to-PEV mapping changes with learning and development.

Despite these limitations, this demonstration establishes that the core mechanism proposed by PEP, that behavior reflects optimization of a learnable positive experience function, can be empirically investigated using current AI technologies. The above-chance prediction accuracy, though modest, provides initial evidence that the PEP framework captures genuine aspects of how conscious systems navigate their experiential state space.

It should be noted that the PEP is designed to complement rather than replace existing models of reinforcement, motivation, and consciousness. It provides a higher-order organizing principle that can unify diverse findings across fields while preserving the explanatory value of mechanistic models.

In summary, The PEP establishes a foundation for quantitative prediction in consciousness research. Unlike purely descriptive theories, it generates specific hypotheses about behavioral outcomes based on the relative PEV values of available states. The framework suggests that AI embeddings could potentially serve as proxies for UCC configurations, opening concrete avenues for empirical validation. This transforms the theory from a philosophical proposition into a falsifiable scientific hypothesis with measurable predictions.

## 6. Methods

### 6.1. Data Collection

We began with a publicly available dataset of 416,809 English tweets labeled with emotion categories from the Twitter Emotion Classification Dataset on Kaggle [18]. To ensure sufficient context, we filtered for tweets containing at least 30 words, yielding approximately 39,624 candidates (9.5% retention).

From these, we used the Qwen/Qwen3-Next-80B-A3B-Instruct large language model via DeepInfra API (https://deepinfra.com/) to identify tweets describing binary choices between experiential alternatives. The extraction process specifically targeted sensory and perceptual experiences rather than abstract concepts, asking the model to identify concrete experiences like "hearing loud music," "feeling warmth," or "seeing bright lights". The prompt below was used to perform extraction:

*CHOICE_EXTRACTION_PROMPT = """*

*# Twitter Text Analysis Prompt for Choice Extraction*

*## Task*

*Analyze the following Twitter text to identify if it describes a situation where the author faced a choice between two distinct experiences and made a decision.*

*## Instructions*

*1. Determine if the text contains a choice between two experiential options (not just abstract concepts)*

*2. If yes, extract the **perceived experiences** associated with each choice option*

3. Identify which choice was made

4. Remember: experiences include not just immediate actions, but associated feelings, sensations, outcomes, consequences, and contextual perceptions

## Important Notes

- Look for implicit choices, not just explicit "should I do X or Y" statements

- Each choice can involve multiple associated **concrete perceptual experiences** (e.g., "going to the dance" includes: hearing rhythmic music, seeing moving bodies, feeling bodily movement, sensing warmth of crowded space, etc.)

- Extract **specific sensory and perceptual experiences** that describe what the person would see, hear, feel, taste, smell, or sense in their body

- Valid experiences:

  - Sensory perceptions: "hearing loud music", "seeing bright lights", "tasting sweet chocolate"

  - Bodily sensations: "feeling muscle exertion", "sensing warmth on skin", "feeling heart racing"

  - Physical environmental perceptions: "perceiving crowded space", "sensing cold air"

- Avoid:

  - Abstract emotions without sensory grounding: "excitement", "happiness", "satisfaction"

  - Vague conceptual states: "relaxation", "accomplishment", "connection"

  - Social/cognitive concepts: "social atmosphere", "sense of achievement"

- The choice made may be stated directly or implied through context

## Response Format

Return ONLY valid JSON with this exact structure:

```json

{{

  "contain_two_choices_between_experiences": boolean,

  "text_of_experiences_for_choice_one": ["experience1", "experience2", ...],

  "text_of_experiences_for_choice_two": ["experience1", "experience2", ...],
```

*"choice_made": "one" | "two" | null*

*}}*

```

*## Twitter Text to Analyze*

*{tweet_text}*

*Remember: Return ONLY the JSON object, no additional text or explanation.*

*"""*

The model also determined which choice the person actually made. After filtering for cases with clear choice indicators and valid experience descriptions for both options, we obtained 1,187 usable choice pairs. Each pair consisted of two sets of experience descriptions (mean ≈ 4.09 experiences for chosen options; ≈ 3.98 for unchosen) along with an indicator of which option was chosen.

### 6.2. Experience Representation

We represented each experience description using dense vector embeddings from the BAAI/bge-small-en-v1.5 model (384 dimensions) via Hugging Face (https://huggingface.co/) Inference API. Unlike conventional pooling approaches, we did not average embeddings for multiple experiences within a choice option. Instead, consistent with the Positive Experience Principle, we learned PEV at the single-experience level and aggregated to the choice level by summation:

$$\text{total-PEV(choice)} = \sum_{i=1}^{N} \text{PEV}(x_i)$$

where $x_i$ is the embedding of experience i. Since reliable intensity labels were not available, for simplicity, we used unweighted summation (all weights = 1). This formulation naturally handles different numbers of experiences per option and directly implements the PEP additivity assumption.

To improve computational efficiency and model stability, we applied standardization (z-score normalization) followed by Principal Component Analysis, reducing the 384 dimensions to 128 while retaining 85-90% of the variance. All preprocessing parameters (standardization statistics and PCA transformation) were fit exclusively on training data to prevent information leakage.

### 6.3. Model Architecture and Training

We implemented a linear model that maps a single experience embedding to a scalar PEV value: $\text{PEV}(x) = w^T x + b$. An optional bounded output via tanh with scale s could be applied, though the default configuration used unbounded outputs.
```

The model was trained using a pairwise logistic ranking loss (Bayesian Personalized Ranking) that directly optimizes the constraint that chosen experiences should receive higher total PEV than unchosen alternatives:

$$L_{BPR} = -\log \sigma\left(\frac{\Delta PEV}{\tau}\right)$$

where ΔPEV = PEV_chosen - PEV_unchosen and τ is a temperature parameter. We additionally employed in-batch negatives: with batch size B = 32, all unchosen totals in the batch served as negatives for each chosen total.

We used 4-fold cross-validation, training separate models on approximately 75% of the data and evaluating on the held-out 25%. Training employed the Adam optimizer (learning rate = $10^{-3}$) with multiple regularization techniques to prevent overfitting: dropout (p = 0.5), weight decay ($10^{-2}$), PEV L2 penalty ($10^{-3}$), and optional L1 penalties and input noise. Early stopping halted training if validation ranking loss did not improve for 5 consecutive epochs. Learning rate was reduced by 0.5 if validation loss plateaued for 3 epochs. Training typically converged within 50 epochs maximum. Gradient clipping (max norm = 1.0) was applied for stability.

### 6.4. Evaluation

The primary evaluation metric was pairwise accuracy: the proportion of validation cases where the model correctly assigned higher total PEV to the chosen experience (i.e., PEV_chosen > PEV_unchosen). Additional metrics included validation ranking loss, mean PEV margin (ΔPEV), pass rate (proportion with ΔPEV > 1.0), and average choice probability σ(ΔPEV/τ). Statistical significance was assessed using a one-sample t-test comparing mean pairwise accuracy across the 4 folds to chance level (50%).

All analyses were conducted in Python 3.8+ using PyTorch 1.13+ for model training, scikit-learn 1.0+ for preprocessing (StandardScaler, PCA, KFold), and standard scientific packages (NumPy, Matplotlib, Seaborn). All code and processing scripts are available at https://github.com/suzheng/pep-choice-pred.